\documentclass[12pt,a4paper]{JHEP3}
\usepackage{amssymb,amsmath,amsfonts,cite}

\newcommand{\Nmax}{\ensuremath{N_{\mathrm{max}}}}
\newcommand{\vev}[1]{\ensuremath{\langle #1\rangle}}

\title{A Holographic Holographic Bound and the Black Hole S-Matrix}

\author{{Michael Gary}\\
           Institute for Theoretical Physics\\
           Vienna University of Technology\\
           Wiedner Hauptstr. 8--10/136\\
           A-1040 Vienna, Austria\\
           Email: \email{mgary@hep.itp.tuwien.ac.at}}

\abstract{
  Holographic bounds have been derived using explicitly gravitational arguments. Motivated by explicit constructions of bulk wavepackets from observables in the boundary CFT, we derive a holographic bound in the context of the gauge/gravity correspondence within the dual field theory. We verify the consistency of the bound with the program of determining the Black Hole S-Matrix from the AdS/CFT correspondence.
}


\preprint{TUW--12--24}

\begin{document}

\section{Introduction}\label{intro}

The physics of black holes provides a unique glimpse into the framework of quantum gravity. In attempting to understand the dynamics of quantum black holes, we encounter profound difficulties and conflicts with our usual Quantum Field Theory intuition. In particular, without some new physical principle, the logic of local Quantum Field Theory in curved spacetime leads to a contradiction---black hole evaporation \cite{Hawking:1974sw} forces us to sacrifice one (or more) assumptions when considering gravitational systems: locality, unitarity, or a finite number of low energy degrees of freedom\footnote{Typically, when considering routes out of the black hole information problem, this third option is introduced in the form of black hole remnants, which must have an arbitrarily large degeneracy of internal degrees of freedom. Often, such a degeneracy leads to production instabilities\cite{Giddings:1993km}.}. 

Motivated by the fact that the entropy of a black hole scales as its horizon area (in Planck units), rather than its volume \cite{Bekenstein:1973ur}, Bekenstein conjectured a ``Holographic Bound'' on the number of localized degrees of freedom in a gravitational system \cite{Bekenstein:1980jp}. The holographic bound in quantum gravity seems to be analogous to the Heisenberg uncertainty relation in quantum mechanics. Much as the uncertainty relation can be derived from a more fundamental formulation of quantum mechanics, 't Hooft and Susskind have proposed extending the holographic bound to a more general ``Holographic Principle''---the number of degrees of freedom of a quantum gravitational system is determined by the area enclosing the system\cite{'tHooft:1993gx,Susskind:1994vu}. Such a principle preserves unitarity and seems to be well behaved in the infra-red while apparently subtlely sacrificing locality. Subsequently, the holgraphic bound has been extended with a more covariant formulation \cite{Bousso:1999xy}, and an excellent review of holographic bounds and evidence in support thereof can be found in \cite{Bousso:2002ju}. 

While general principles are important for formulating theories, without specific examples, it is often difficult to develop physical intuition. It is therefore extremely fortuitous that the AdS/CFT correspondence \cite{Maldacena:1997re} exists, as it is a specific manifestation of the general principle of holography. Bulk graviational degrees of freedom are described by a non-gravitational theory living on the boundary, which acts as a holographic screen. As such, it should be expected that the usual holographic bounds, derived from bulk gravitational reasoning, should be visible from the boundary, non-gravitational, perspective. While the theory is manifestly holographic and thus satisfies a holographic bound for the spacetime as a whole \cite{Susskind:1998dq}, it should in principle be possible to place holographic bounds on subregions of the bulk in addition to the spacetime in its entirety. While some progress has been made towards deriving bulk holographic bounds from within the field theory \cite{Hamilton:2006az}, it is clear that much work remains. As a specific example of such a bulk holographic bound, one should be able to bound the number of degrees of freedom accessible within a single approximately flat AdS region via a purely boundary argument. We provide such an argument and derive a bound on the number of localized degrees of freedom. 

In the remainder of the paper, we will motivate and explain our construction for single particle states in section \ref{dCompact}, which we will generalize to multi-particle states in section \ref{multiParticle}, leading to a holographic bound. We will then discuss a specific application and test of this holographic bound to the case of small black holes in the bulk of AdS in section \ref{BHSMatrix}. Holographic bounds are often motivated by considering black holes. Here, we wish to reverse this logic, deriving a holographic bound without reference to a specific black hole system, then testing the bound by a comparison to the properties of black holes. Since black holes saturate the usual holographic bound, they provide an ideal testing ground for conjectured new holographic bounds. Finally, we will conclude with some remarks in section \ref{conclusions}.

\section{Boundary-Compact Sources and Single Particle States}\label{dCompact}

The states we will consider were introduced for the purpose of using the AdS/CFT correspondence to determine the flat-space gravitational S-Matrix. In particular, as proposed in \cite{JoeSMatrix,Susskind:1998vk}, we consider wavepackets which remain localized within a single AdS volume and take the large $R$ limit of AdS. Much progress has been made in recent years in this program \cite{MikeSteveJoao,SMatrix,Constraints,Heemskerk:2009pn,Fitzpatrick:2011jn}, and we will take advantage of some of the tools developed in the course of this work. 

In the context of scattering localized wavepackets to recover the flat-space S-Matrix, it is important that the wavepackets not interact strongly near the boundary of AdS, as such interactions would spoil the isolation of a single, well-localized scattering event \cite{Giddings:1999jq,MikeSteveJoao,SMatrix,Constraints}. To avoid such near-boundary interactions, it is sufficient to consider boundary sources with non-overlapping compact support. A particular class of such ``boundary-compact'' sources were introduced in \cite{MikeSteveJoao} and shown to produce wavepackets approximately localized within a region of spatial size $L\ll R$ in the bulk of AdS. Such sources have the property that they have compact support on the boundary $S^{d-1}$ with angular width $\delta\theta$ as well as compact support in time. 

In general, to have wavepackets which are localized in a region small as compared to the AdS radius, we require
\begin{equation}
\delta\theta \gg \frac{1}{\omega R}\ .
\end{equation}

One method to ensure a large separation between the wavepacket localization size $L$ and the AdS size $R$ is to ensure that they scale with different powers of a fiducial scaling parameter, $\eta$, as advocated in \cite{JoeSMatrix,MikeSteveJoao}. In particular, if we choose the scalings
\begin{align}\label{JoeJoaoScaling}
R &\sim \eta R_0 & L &\sim \sqrt{\eta} L_0\ ,
\end{align}
we find 
\begin{equation}
\delta\theta \sim \frac{1}{\omega \sqrt{L_0 R}} \gg \frac{1}{\omega R}\ .
\end{equation}
However, for our purposes, this additional separation of scales will not be of particular importance. We will therefore take $L=R$ unless otherwise explicitly noted. 

There is an important caveat when using boundary-compact wavepackets---the tails are power-law rather than Schwarz \cite{SMatrix}, as holds for ``regular wavepackets'' \cite{Reed:1979ne} typically used in rigorous treatments of flat space scattering. Due to the nature of AdS space, any wavepacket created from a spatially compact boundary source will have such power-law tails. While boundary-compact wavepackets are sharply peaked, for masses too far above the Breitenlohner-Freedman bound \cite{Breitenlohner:1982}, there is a danger that the tail wags the dog, where the norm of the state is actually dominated by the tail of the wavepacket. 

\section{Multi-Particle States and the Holographic Bound}\label{multiParticle}

Consider creating a multi-particle state using non-overlapping boundary-compact wavepackets. From the bulk perspective, it is necessary that the sources be placed in a non-overlapping manner on the boundary $S^{d-1}$ to avoid large interactions in the near-boundary region.\footnote{In the special case of pure gravity this argument does not apply, as the interaction strength is governed by the bulk stress tensor, which redshifts to zero near the boundary, while the norm of the graviton field remains constant (like a scalar with $\Delta=d$) \cite{Constraints}. However, as soon as there are other interacting scalars present, such as in the consistent truncation of IIB to AdS$_5$, near-boundary interactions again become problematic.} In fact, it is also possible to see this requirement purely from within the CFT. In particular, turning on overlapping sources in the CFT introduces divergences, which we should avoid in order to have well-defined, finite correlators. If we consider working at finite but large AdS radius $R$, these wavepackets will have a finite angular width $\delta\theta$, meaning that there is a maximum bulk particle number that can be achieved with this construction, since there is a maximum number $\Nmax$ of non-overlapping sources that can be placed on the boundary. 

A straightforward counting argument lets us determine an upper bound on $\Nmax$ by comparing the boundary volume of a single source $V_{\mathrm{source}}$ to the total boundary volume, 
\begin{equation}
\Nmax \leq \frac{V_{S^{d-1}}}{V_{\mathrm{source}}}\ .
\end{equation}
The problem of determining $\Nmax$ now reduces to determining the boundary spatial volume occupied by a single source. If we take the sources to be spherically symmetric, as in \cite{MikeSteveJoao}, $V_{\mathrm{source}}$ is given by the volume of a hyperspherical cap of polar angular width $\delta\theta$. Thus, we find
\begin{equation}
\Nmax \leq \frac{2}{I_{\sin^2\delta\theta}\left(\frac{d-1}{2},\frac{1}{2}\right)}\ ,
\end{equation}
where $I_x(a,b)$ is the regularized incomplete beta function. Taking the flat-space limit of large AdS radius $R$ at fixed scattering energy $\omega$ is equivalent to considering the case $\delta\theta\ll1$, in which case we find
\begin{equation}\label{CFTBound}
\Nmax \sim \frac{(d-1)B\left(\frac{d-1}{2},\frac{1}{2}\right)}{(\delta\theta)^{d-1}} \stackrel{\ll}{\sim} (d-1)B\left(\frac{d-1}{2},\frac{1}{2}\right)(\omega R)^{d-1}\ .
\end{equation}
Comparing this to the covariant entropy bound \cite{Bousso:1999xy}
\begin{equation}
S\leq \frac{A}{4} = \frac{\pi^{d/2}}{2\Gamma(d/2)}R^{d-1}\ ,
\end{equation}
we find that the bound derived from the dual field theory differs by a factor of $\omega^{d-1}$ (the bounds coincide for energies of order the Planck scale). Such a difference may be accounted for by the fact that $\Nmax$ must be much less than $(\omega R)^{d-1}$; however, this is unlcear from the present analysis. This factor of $\omega^{d-1}$ is particular puzzling when considering the high energy limit $\omega\gg\frac{1}{\ell_p}$. In this limit, the upper bound on $\Nmax$ diverges. While this does not directly imply that it is possible to excite an infinite number of high energy modes in a single AdS volume, our argument bounding the number of states no longer holds and some additional argument must come into play if our usual gravitational intuition is to remain valid. This is in direct contrast to usual holographic bounds. In particular, for asymptotically high energies, we ought to expect the generic multi-particle state to form a black hole. 

If we are to recover the usual holographic behavior, some additional argument must come into play. One such restriction on the number of states in a single AdS region may come from considering the unusual power-law tails associated with boundary-compact wavepackets \cite{MikeSteveJoao,SMatrix,Constraints}. In particular, since boundary-compact states are not as well-localized in the bulk as the usual ``regular'' wavepackets considered in flat-space scattering, enough of the norm of the state may be outside the region of size $R$ to evade arguments about black-hole formation, or at a minimum arguments about the formation of a black hole within a single region of size $R$. 

It is also interesting to consider the particular case where we take the scaling limit (\ref{JoeJoaoScaling}). In such a limit, the holographic bound (\ref{CFTBound}) coming from the conformal field theory reduces to
\begin{equation}
\Nmax \sim (d-)B\left(\frac{d-1}{2},\frac{1}{2}\right)\left(\omega\sqrt{L_0 R}\right)^{d-1}\ .
\end{equation}
Such a scaling limit seems to severely limit the number of states accessible within the approximately flat region relative to the bound \ref{CFTBound}. However, again, the puzzling $\omega$ dependence remains unchanged.

\section{The Black Hole S-Matrix}\label{BHSMatrix}

We can gain further insight into the holographic bound (\ref{CFTBound}) by considering the time reversed situation---rather than creating a multi-particle state in the bulk from insertion of boundary operators, imagine measuring a bulk multi-particle state by ``instrumenting'' the boundary with operators. The highest entropy localized state in the bulk should correspond to a black hole, which, if allowed to decay, will result in a number of Hawking particles $N$. If the lifetime of the black hole is short as compared to the AdS time $R$, the particles should all arrive at the boundary at approximately the same time. Assuming the boundary CFT knows about the flat space S-Matrix, and, in particular, the black hole S-Matrix \cite{Veneziano:2004er,Giddings:2007qq,Giddings:2009gj,Giddings:2011xs}, we must be able to detect all of these particles in the CFT, and in particular we must find $N\leq\Nmax$. 

The lifetime of a black hole scales as\footnote{We are effectively enforcing outgoing boundary conditions at the boundary of AdS, so black holes will evaporate rather than coming into equilibrium with their Hawking radiation, as would happen for a large black hole in AdS with reflecting boundary conditions.}
\begin{equation}\label{BHLifetime}
T_{\mathrm{evap}} \sim R_H S \sim \frac{R_H^d}{\ell_p^{d-1}}\ ,
\end{equation}
while the typical number of Hawking quanta emitted by a black hole is given by its entropy
\begin{equation}\label{NumHawk}
\vev{N} \sim S \sim \left(\frac{R_H}{\ell_p}\right)^{d-1}\ .
\end{equation}
A black hole with a lifetime of order the AdS time\footnote{Here we will consider black holes in pure AdS$_{d+1}$. It has been argued that small black holes in AdS$\times M$ for some compact $M$ are unstable to localization in $M$, leading to effectively 10/11 dimensional black holes \cite{Gregory:1993vy,Horowitz:1999uv}. For a D-dimensional black hole with an evaporation time of order $R$, the typical number of expected Hawking quanta is $\vev{N}\sim\left(\frac{R}{\ell_p}\right)^{(D-2)/(D-1)}$. For $d>2$, $\Nmax > \vev{N}$ for any number of compact dimensions.} can be expected to emit
\begin{equation}
\vev{N} \sim \left(\frac{R}{\ell_p}\right)^{\frac{d-1}{d}}
\end{equation}
Hawking quanta. 

As we have already noted, it is possible to instrument the boundary of AdS with more high energy detectors than with low energy detectors. Therefore, if we can detect sufficient quanta with low energy detectors, we can certainly detect sufficient quanta with smaller, higher energy detectors. This, in combination with the fact that smaller black holes emit fewer, higher energy Hawking quanta, means that if we can instrument the boundary sufficiently well to detect all Hawking quanta from a black hole with evaporation time scale of order the AdS time, we can certainly detect all Hawking quanta from smaller black holes. For simplicity, we may take the Hawking quanta to all be emitted at approximately the Hawking temperature
\begin{equation}
\vev{\omega_H} \sim \frac{\kappa}{2\pi} \sim \frac{1}{R_H}\ .
\end{equation}
For a black hole with a lifetime of order the AdS time, the average energy is
\begin{equation}
\vev{\omega} \sim \frac{1}{(R\ell_p^{d-1})^{1/d}}\ ,
\end{equation}
and we can detect at most
\begin{equation}
\Nmax \sim \left(\frac{R}{\ell_p}\right)^{\frac{(d-1)^2}{d}}
\end{equation}
quanta of this energy. For $R\gg\ell_p$, $\Nmax \gg \vev{N}$ the expected number of Hawking quanta, and thus it is possible to detect all Hawking quanta of black holes with lifetimes short as compared to the AdS time.

\section{Conclusions}\label{conclusions}

In summary, we have derived a holographic bound from purely field theoretic considerations in the context of the Gauge/Gravity duality. While there are some limitations to our argument---in particular, the restriction on particle number does not apply in the special case of pure gravity, since it is possible to have overlapping boundary sources without introducing divergences\cite{Constraints}---near-boundary divergences due to overlapping sources are generic in AdS/CFT, and thus our argument is quite general. More startlingly, in certain scaling regimes which seem to satisfy all requirements from the boundary perspective, it seems possible to localize an arbitrarily large number of high energy states within a single AdS region. Unless effects such as those discussed in \cite{SMatrix,Constraints} rule out such states, this scaling limit seems to contradict our usual notions about entropy bounds and black hole formation. At a very minimum, the boundary construction leads to a severe overcounting of bulk states.

While we have derived the holographic bound by thinking about constructing a multi-particle state using boundary operators, the time-reverse of this picture corresponds to an outgoing state with some maximal number of particles $\Nmax$. It is this particular version of the bound that places a limit on the size of black holes the S-Matrix captures, as the number of out-going Hawking quanta is expected to be of order
\begin{equation}
N \sim \frac{A}{4}\ .
\end{equation}
Our holographic bound is consistent with the notion that only states which are short-lived as compared to the AdS time $R$ should be captured by the flat space limit and can be considered as supporting the notion that the flat space graviational S-Matrix can be determined from AdS/CFT. 

The holographic principle in general, and the AdS/CFT correspondence in particular, seem to point towards giving up locality in order to resolve the black hole information problem. The CFT, which is conjectured to provide an exact dual description of the bulk gravitation theory, manifestly evolves unitarily, and strongly suggests the unitary evolution can be extended to the bulk\cite{Marolf:2008mf,Marolf:2008mg,Heemskerk:2012mn}. Furthermore, the AdS/CFT correspondence can be argued to be inconsistent with the idea of black hole remanants---the density of states of in the CFT on $S^d\times\mathbb{R}$ scales linearly with $\omega$ at low energies, which is inconsistent with the large degeneracy of low energy states corresponding to black hole remnants\footnote{This argument is due to A. Strominger and was conveyed to me by D. Harlow.}. 

While the AdS/CFT correspondence is manifestly holographic, the exact nature of the non-locality which leads to the resolution of the black hole information paradox remains unclear. One avenue towards understanding where locality breaks down is to approach saturation of the holographic bound. Much like coherent states, which saturate the uncertainty bound in quantum mechanics, are maximally classical, states which saturate the holographic bound should in some sense be maximally local, and thus provide an excellent arena for exploring the breakdown of locality in quantum gravity. Boundary-compact wavepackets seem to be an excellent candidate for a class of states to use to saturate holographic bounds in AdS/CFT, and while the counting argument we provide seems to allow for oversaturation of the holographic bound, it seems plausible that a more careful treatment of issues such as wavepacket tails should correct for this overcounting.

\acknowledgments

I would like to thank Steve Giddings for discussions, especially regarding aspects of the Black Hole S-Matrix, Daniel Harlow for enlightening correspondence on the topic of black hole remnants, especially in the context of the AdS/CFT correspondence, and Daniel Grumiller for remarks as the paper was in preparation. This work was supported by the START project Y435-N16 of the Austrian Science Fund (FWF).

\providecommand{\href}[2]{#2}\begingroup\raggedright\endgroup


\begin{thebibliography}{50}

\bibitem{Hawking:1974sw}
  S.~W.~Hawking,
  ``Particle Creation by Black Holes,''
  Commun.\ Math.\ Phys.\  {\bf 43} (1975) 199
   [Erratum-ibid.\  {\bf 46} (1976) 206].

\bibitem{Giddings:1993km}
  S.~B.~Giddings,
  ``Constraints on black hole remnants,''
  Phys.\ Rev.\ D {\bf 49} (1994) 947
  \href{http://www.arXiv.org/abs/hep-th/9304027}{{\tt[hep-th/9304027]}}.

\bibitem{Bekenstein:1973ur}
  J.~D.~Bekenstein,
  ``Black holes and entropy,''
  Phys.\ Rev.\ D {\bf 7} (1973) 2333.

\bibitem{Bekenstein:1980jp}
  J.~D.~Bekenstein,
  ``A Universal Upper Bound on the Entropy to Energy Ratio for Bounded Systems,''
  Phys.\ Rev.\ D {\bf 23} (1981) 287.

\bibitem{'tHooft:1993gx}
  G.~'t Hooft,
  ``Dimensional reduction in quantum gravity,''
  \href{http://www.arXiv.org/abs/hep-th/9310026}{{\tt gr-qc/9310026}}.

\bibitem{Susskind:1994vu}
  L.~Susskind,
  ``The World as a hologram,''
  J.\ Math.\ Phys.\  {\bf 36} (1995) 6377
  \href{http://www.arXiv.org/abs/hep-th/9409089}{{\tt[hep-th/9409089]}}.

\bibitem{Bousso:1999xy}
  R.~Bousso,
  ``A Covariant entropy conjecture,''
  JHEP {\bf 9907} (1999) 004
  \href{http://www.arXiv.org/abs/hep-th/9905177}{{\tt[hep-th/9905177]}}.

\bibitem{Bousso:2002ju}
  R.~Bousso,
  ``The Holographic principle,''
  Rev.\ Mod.\ Phys.\  {\bf 74} (2002) 825
  \href{http://www.arXiv.org/abs/hep-th/0203101}{{\tt[hep-th/0203101]}}.

\bibitem{Maldacena:1997re}
  J.~M. Maldacena, 
  ``The large N limit of superconformal field theories and supergravity,''
  {\em Adv. Theor. Math. Phys.} {\bf 2} (1998) 231--252,
  \href{http://www.arXiv.org/abs/hep-th/9711200}{{\tt[hep-th/9711200]}}.

\bibitem{Susskind:1998dq}
  L.~Susskind and E.~Witten,
  ``The Holographic bound in anti-de Sitter space,''
  \href{http://www.arXiv.org/abs/hep-th/9805114}{{\tt hep-th/9805114}}.

\bibitem{Hamilton:2006az}
  A.~Hamilton, D.~N.~Kabat, G.~Lifschytz and D.~A.~Lowe,
  ``Holographic representation of local bulk operators,''
  Phys.\ Rev.\ D {\bf 74} (2006) 066009
  \href{http://www.arXiv.org/abs/hep-th/0606141}{{\tt [hep-th/0606141]}}.

\bibitem{JoeSMatrix}
  J.~Polchinski,
  ``S matrices from AdS space-time,''
  \href{http://www.arXiv.org/abs/hep-th/9901076}{{\tt hep-th/9901076}}.

\bibitem{Susskind:1998vk}
  L.~Susskind,
  ``Holography in the flat space limit,''
  \href{http://www.arXiv.org/abs/hep-th/9901079}{{\tt hep-th/9901079}}.

\bibitem{MikeSteveJoao}
  M.~Gary, S.~B.~Giddings and J.~Penedones,
  ``Local bulk S-matrix elements and CFT singularities,''
  Phys.\ Rev.\ D {\bf 80} (2009) 085005
  \href{http://www.arXiv.org/abs/0903.4437}{{\tt[arXiv:0903.4437 [hep-th]]}}.

\bibitem{SMatrix}
  M.~Gary and S.~B.~Giddings,
  ``The Flat space S-matrix from the AdS/CFT correspondence?,''
  Phys.\ Rev.\ D {\bf 80} (2009) 046008
  \href{http://www.arXiv.org/abs/0904.3544}{{\tt[arXiv:0904.3544 [hep-th]]}}.

\bibitem{Constraints}
  M.~Gary and S.~B.~Giddings,
  ``Constraints on a fine-grained AdS/CFT correspondence,''
  \href{http://www.arXiv.org/abs/1106.3553}{{\tt arXiv:1106.3553 [hep-th]}}.

\bibitem{Heemskerk:2009pn}
  I.~Heemskerk, J.~Penedones, J.~Polchinski and J.~Sully,
  ``Holography from Conformal Field Theory,''
  JHEP {\bf 0910} (2009) 079
  \href{http://www.arXiv.org/abs/0907.0151}{{\tt[arXiv:0907.0151 [hep-th]]}}.\\
  T.~Okuda and J.~Penedones,
  ``String scattering in flat space and a scaling limit of Yang-Mills correlators,''
  Phys.\ Rev.\ D {\bf 83} (2011) 086001
  \href{http://www.arXiv.org/abs/1002.2641}{{\tt[arXiv:1002.2641 [hep-th]]}}.\\
  I.~Heemskerk and J.~Sully,
  ``More Holography from Conformal Field Theory,''
  JHEP {\bf 1009} (2010) 099
  \href{http://www.arXiv.org/abs/1006.0976}{{\tt[arXiv:1006.0976 [hep-th]]}}.\\
  A.~L.~Fitzpatrick, E.~Katz, D.~Poland and D.~Simmons-Duffin,
  ``Effective Conformal Theory and the Flat-Space Limit of AdS,''
  JHEP {\bf 1107} (2011) 023
  \href{http://www.arXiv.org/abs/1007.2412}{{\tt[arXiv:1007.2412 [hep-th]]}}.\\
  A.~L.~Fitzpatrick, J.~Kaplan, J.~Penedones, S.~Raju and B.~C.~van Rees,
  ``A Natural Language for AdS/CFT Correlators,''
  JHEP {\bf 1111} (2011) 095
  \href{http://www.arXiv.org/abs/1107.1499}{{\tt[arXiv:1107.1499 [hep-th]]}}.\\
  A.~L.~Fitzpatrick and J.~Kaplan,
  ``Analyticity and the Holographic S-Matrix,''
  \href{http://www.arXiv.org/abs/1111.6972}{{\tt arXiv:1111.6972 [hep-th]}}.\\
  D.~Nandan, A.~Volovich and C.~Wen,
  ``On Feynman Rules for Mellin Amplitudes in AdS/CFT,''
  JHEP {\bf 1205} (2012) 129
  \href{http://www.arXiv.org/abs/1112.0305}{{\tt [arXiv:1112.0305 [hep-th]]}}.\\
  A.~L.~Fitzpatrick and J.~Kaplan,
  ``Unitarity and the Holographic S-Matrix,''
  \href{http://www.arXiv.org/abs/1112.4845}{{\tt arXiv:1112.4845 [hep-th]}}.

\bibitem{Fitzpatrick:2011jn}
  A.~L.~Fitzpatrick and J.~Kaplan,
  ``Scattering States in AdS/CFT,''
  \href{http://www.arXiv.org/abs/1104.2597}{{\tt arXiv:1104.2597 [hep-th]}}.

\bibitem{Giddings:1999jq}
  S.~B.~Giddings,
  ``Flat space scattering and bulk locality in the AdS / CFT correspondence,''
  Phys.\ Rev.\ D {\bf 61} (2000) 106008
  \href{http://www.arXiv.org/abs/hep-th/9907129}{{\tt[hep-th/9907129]}}.

\bibitem{Reed:1979ne}
  M.~Reed and B.~Simon,
  ``Methods Of Mathematical Physics. Vol. 3: Scattering Theory,''
  New York, Usa: Academic ( 1979) 463p

\bibitem{Breitenlohner:1982}
  P.~Breitenlohner and D.~Z.~Freedman,
  ``Positive Energy in anti-De Sitter Backgrounds and Gauged Extended Supergravity,''
  Phys.\ Lett.\ B {\bf 115} (1982) 197.\\
  P.~Breitenlohner and D.~Z.~Freedman,
  ``Stability in Gauged Extended Supergravity,''
  Annals Phys.\  {\bf 144} (1982) 249.

\bibitem{Giddings:2011xs}
  S.~B.~Giddings,
  ``The gravitational S-matrix: Erice lectures,''
  \href{http://www.arXiv.org/abs/1105.2036}{{\tt arXiv:1105.2036 [hep-th]}}.

\bibitem{Giddings:2011dr}
  S.~B.~Giddings,
  ``Is string theory a theory of quantum gravity?,''
  \href{http://www.arXiv.org/abs/1105.6359}{{\tt arXiv:1105.6359 [hep-th]}}.

\bibitem{Veneziano:2004er}
  G.~Veneziano,
  ``String-theoretic unitary S-matrix at the threshold of black-hole production,''
  JHEP {\bf 0411} (2004) 001
  \href{http://www.arXiv.org/abs/hep-th/0410166}{{\tt [hep-th/0410166]}}.

\bibitem{Giddings:2007qq}
  S.~B.~Giddings and M.~Srednicki,
  ``High-energy gravitational scattering and black hole resonances,''
  Phys.\ Rev.\ D {\bf 77} (2008) 085025
  \href{http://www.arXiv.org/abs/0711.5012}{{\tt [arXiv:0711.5012 [hep-th]]}}.

\bibitem{Giddings:2009gj}
  S.~B.~Giddings and R.~A.~Porto,
  ``The Gravitational S-matrix,''
  Phys.\ Rev.\ D {\bf 81} (2010) 025002
  \href{http://www.arXiv.org/abs/0908.0004}{{\tt [arXiv:0908.0004 [hep-th]]}}.

\bibitem{Gregory:1993vy}
  R.~Gregory and R.~Laflamme,
  ``Black strings and p-branes are unstable,''
  Phys.\ Rev.\ Lett.\  {\bf 70} (1993) 2837
  \href{http://www.arXiv.org/abs/hep-th/93-1-52}{{\tt [hep-th/9301052]}}.

\bibitem{Horowitz:1999uv}
  G.~T.~Horowitz,
  ``Comments on black holes in string theory,''
  Class.\ Quant.\ Grav.\  {\bf 17} (2000) 1107
  \href{http://www.arXiv.org/abs/hep-th/9910082}{{\tt [hep-th/9910082]}}.

\bibitem{Marolf:2008mf}
  D.~Marolf,
  ``Unitarity and Holography in Gravitational Physics,''
  Phys.\ Rev.\ D {\bf 79} (2009) 044010
  \href{http://www.arXiv.org/abs/0808.2842}{{\tt[arXiv:0808.2842 [gr-qc]]}}.

\bibitem{Marolf:2008mg}
  D.~Marolf,
  ``Holographic Thought Experiments,''
  Phys.\ Rev.\ D {\bf 79} (2009) 024029
  \href{http://www.arXiv.org/abs/0808.2845}{{\tt[arXiv:0808.2845 [gr-qc]]}}.

\bibitem{Heemskerk:2012mn}
  I.~Heemskerk, D.~Marolf and J.~Polchinski,
  ``Bulk and Transhorizon Measurements in AdS/CFT,''
  \href{http://www.arXiv.org/abs/1201.3664}{{\tt arXiv:1201.3664 [hep-th]}}.

\end{thebibliography}
\end{document}